\documentclass[twocolumn,showpacs,preprintnumbers,amsmath,amssymb,prb,superscriptaddress,10pt,aps]{revtex4-1}

\usepackage{natbib}
\usepackage{graphicx}
\usepackage{float}
\usepackage{textcomp}
\usepackage{units}
\usepackage{dcolumn}
\usepackage{amsmath,amssymb}
\usepackage{verbatim} 
\usepackage[usenames,dvipsnames]{color}
\usepackage{bm} 

\renewcommand{\figurename}{Figure}

\begin{document}

\title{Tailoring mechanically-tunable strain fields in graphene}

\author{M.~Goldsche}
\thanks{These two authors contributed equally.}
\author{J.~Sonntag}
\thanks{These two authors contributed equally.}
\author{T.~Khodkov}
\affiliation{JARA-FIT and 2nd Institute of Physics, RWTH Aachen University, 52056 Aachen, Germany, EU}
\affiliation{Peter Gr{\"u}nberg Institute (PGI-8/9), Forschungszentrum J{\"u}lich, 52425 J{\"u}lich, Germany, EU}

\author{G.~Verbiest}
\affiliation{JARA-FIT and 2nd Institute of Physics, RWTH Aachen University, 52056 Aachen, Germany, EU}

\author{S.~Reichardt}
\affiliation{JARA-FIT and 2nd Institute of Physics, RWTH Aachen University, 52056 Aachen, Germany, EU}
\affiliation{Physics and Materials Science Research Unit, University of Luxembourg, 1511 Luxembourg, Luxembourg, EU}

\author{C.~Neumann}
\affiliation{JARA-FIT and 2nd Institute of Physics, RWTH Aachen University, 52056 Aachen, Germany, EU}
\affiliation{Peter Gr{\"u}nberg Institute (PGI-8/9), Forschungszentrum J{\"u}lich, 52425 J{\"u}lich, Germany, EU}

\author{T.~Ouaj}
\affiliation{JARA-FIT and 2nd Institute of Physics, RWTH Aachen University, 52056 Aachen, Germany, EU}

\author{N.~von den Driesch}
\author{D.~Buca}
\affiliation{Peter Gr{\"u}nberg Institute (PGI-8/9), Forschungszentrum J{\"u}lich, 52425 J{\"u}lich, Germany, EU}

\author{C.~Stampfer}
\affiliation{JARA-FIT and 2nd Institute of Physics, RWTH Aachen University, 52056 Aachen, Germany, EU}
\affiliation{Peter Gr{\"u}nberg Institute (PGI-8/9), Forschungszentrum J{\"u}lich, 52425 J{\"u}lich, Germany, EU}
\email{Stampfer@physik.rwth-aachen.de}

\begin{abstract}
There are a number of theoretical proposals based on strain engineering of graphene and other two-dimensional materials, however purely mechanical control of strain fields in these
systems has remained a major challenge. The two approaches mostly used so far either couple the electrical and mechanical properties of the system simultaneously or introduce some
unwanted disturbances due to the substrate. Here, we report on silicon micro-machined comb-drive actuators to controllably and reproducibly induce strain in a suspended graphene
sheet, in an entirely mechanical way. We use spatially resolved confocal Raman spectroscopy to quantify the induced strain, and we show that different strain fields can be obtained
by engineering the clamping geometry, including tunable strain gradients of up to 1.4 \%/\textmu m. Our approach also allows for multiple axis straining and is equally applicable to
other two-dimensional materials, opening the door to an investigating their mechanical and electromechanical properties. Our measurements also clearly identify defects at the edges
of a graphene sheet as being weak spots responsible for its mechanical failure.
\end{abstract}

\maketitle

Applying strain fields in graphene not only allows tailoring its mechanical properties \cite{chen2009}, but also reveals fascinating phenomena such as pseudomagnetic fields
\cite{pereira2009b,guinea2010,verbiest2015}, valley filters \cite{fujita2010,grujic2014}, superconductivity \cite{uchoa2013}, or pseudo gravitomagnetic forces \cite{soodchomshom2013}.
Although these phenomena have received much attention theoretically, controlling the required strain fields and strain gradients has remained a major challenge.
Here we present micro-machined comb-drive actuators with integrated graphene for engineering truly mechanically-tunable strain fields.
We use spatially resolved confocal Raman spectroscopy \cite{graf2007} to quantify the strain fields as a function of a controllably induced displacement.
Different strain fields can be obtained by engineering the clamping geometry, including tunable strain gradients up to 1.4 \%/\textmu m.
The presented approach also allows for multiple axis straining and is applicable to the rising number of other two-dimensional materials, thus providing a workhorse for investigating
the fundamental electromechanical properties of 2D materials as well as for developing new sensor and transducer concepts.

Strain is commonly induced in graphene by pulling on suspended sheets with an electrostatic gate \cite{bunch2007,song2011,chen2009,zhang2014,guan2015,nicholl2015} or by bending a
flexible substrate \cite{mohiuddin2009,yoon2011,huang2010,zhang2014,guan2015}.
The obtained strain fields are thus intrinsically linked to either the electronic tuning of the charge carrier density or to the properties of the substrate.
This lack of independent control over strain fields poses a great challenge for any application based on strain engineered graphene.
Moreover, engineering truly controllable local strain patterns in graphene has not been achieved so far.
Our comb-drive (CD) actuators give independent control and allow for engineered strain fields.

\begin{figure*}[!thp]
  \includegraphics[width=178mm]{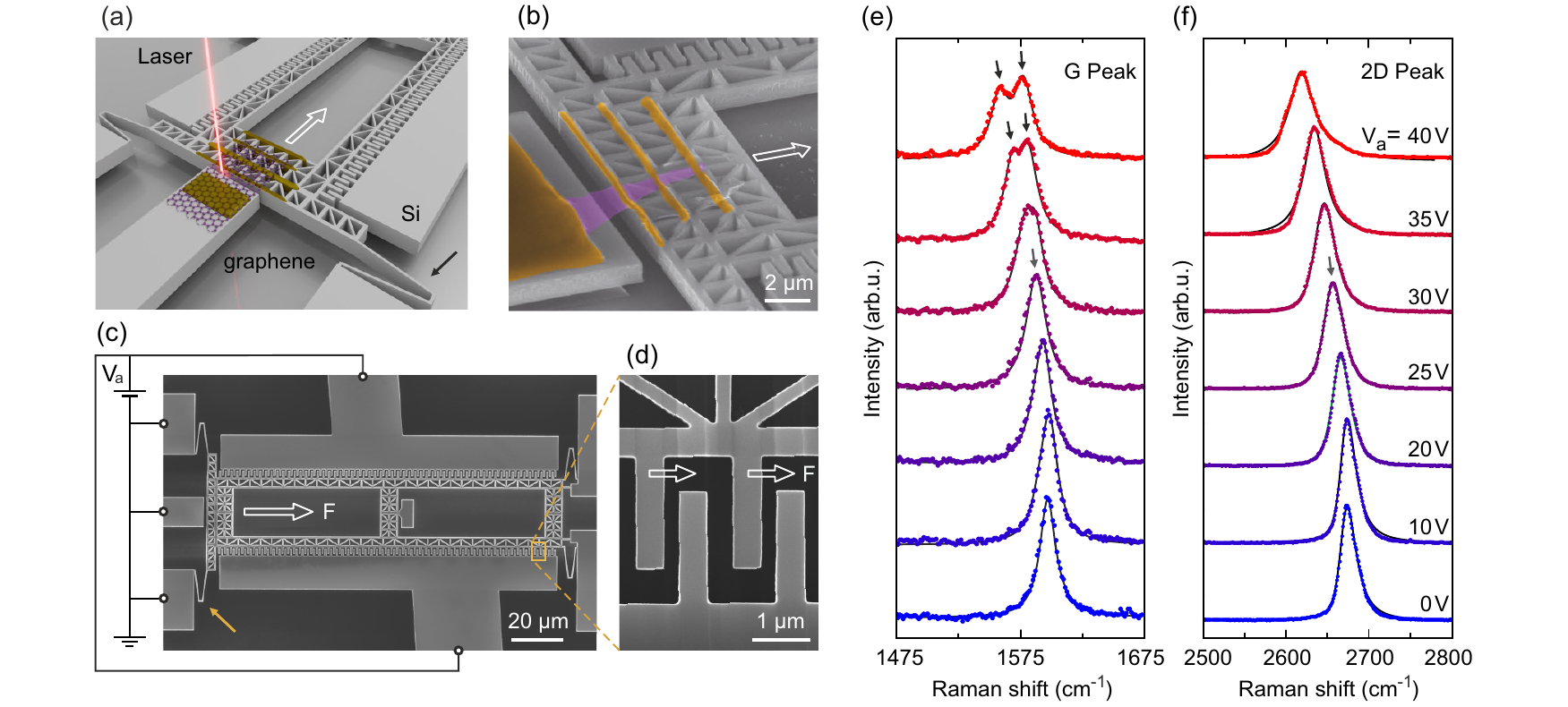}
  \vspace*{-2mm}
  \caption{
    %
    (a) Schematic illustration of a comb-drive (CD) actuator that applies uniaxial strain to the graphene flake by moving into the direction of the white arrow.
    (b) False color scanning electron microscope (SEM) image of a measured device, in which the cross-linked PMMA clamping (yellow) of the graphene (pink) is highlighted.
    (c) Top view SEM image of a fabricated CD actuator.
        The applied potential $V_a$ generates an electrostatic force $F$ between the interdigitated fingers of the fixed and the suspended part of the CD actuator.
        The fixed part is resting on the substrate whereas the suspended part is freely hanging and is held by four V-shaped springs (see yellow arrow).
    (d) Close-up showing the asymmetric distance between the fingers to ensure movement of the suspended part in the direction indicated by the arrow alongside $F$.
    (e) The Raman spectrum of the G peak for increasing $V_a$ shows a red shift and a splitting into two peaks due to strain.
    (f) The Raman spectrum of the 2D peak only shows a red shift with increasing $V_a$.
    The black lines in panels (e) and (f) are fits to Lorentzian peaks that are used to extract $\omega_{\text{G}}$ and $\omega_{\text{2D}}$, respectively.
  }
  \label{fig1}
\end{figure*}

The CD actuators with integrated graphene (schematic in Figure~\ref{fig1}a) are based on surface micro-machining of silicon-on-insulator substrates and on transferring mechanically
exfoliated graphene flakes.
Crucially, we use a polymethylmethacrylat (PMMA) membrane to place the graphene flake onto the CD devices, which is then used to clamp the graphene flake by locally cross-linking it.
After dissolving the remaining PMMA, we use hydrofluoric acid to suspend the complete device (see Methods and Supplementary Figure~1).
Figure~\ref{fig1}b shows a false color scanning electron microscope (SEM) image of such a device.
The actuator consists of a suspended body that is connected by four springs to fixed anchors.
The suspended body has multiple interdigitated fingers with a fixed body (Figure~\ref{fig1}c and d).
A potential difference $V_a$ between the fingers gives rise to an electrostatic force $F =~\eta~V_a^2$ (Figure~\ref{fig1}c), where $2\eta$ is the capacitive coupling \cite{Note1}.
Our devices controllably reach a maximum displacement of 60~nm, which translates into $\sim 6\mathrm{\%}$ of strain in the suspended part of the graphene flake (Supplementary
Figure~3).

\begin{figure*}[!thp]
\includegraphics[width=178mm]{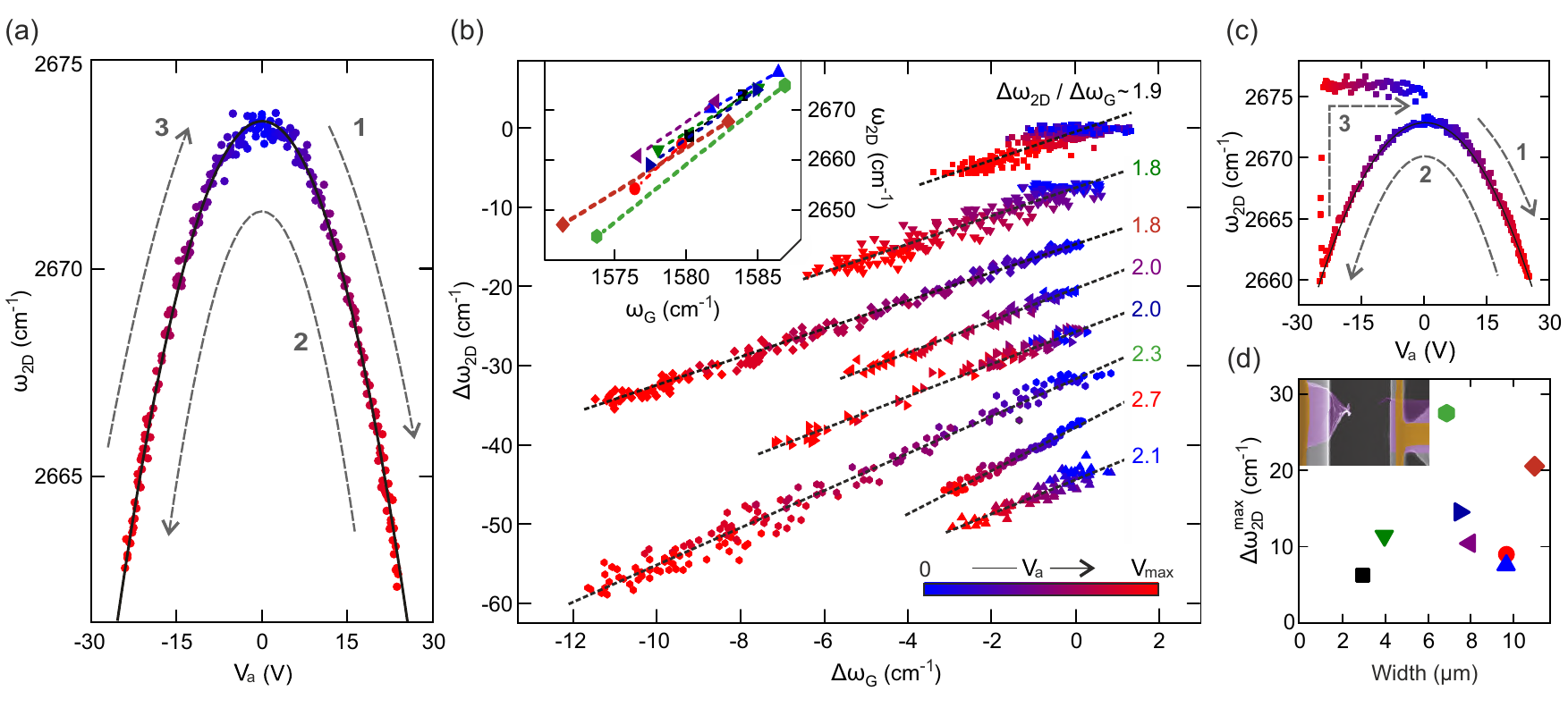}
  \vspace*{-2mm}
  \caption{
    %
    (a) $\omega_{\text{2D}}$ versus $V_a$ shows parabolic behavior. 
        The grey arrows 1 to 3 indicate the sweep direction of the potential $V_a$ in the measurement.
    (b) 
        $\Delta \omega_{\text{2D}}$ versus $\Delta \omega_{\text{G}}$ shows a linear dependence.
        Each curve is offset for clarity by $\Delta\omega_{\text{2D}} = 5$~cm$^{\text -1}$.
        The slope $\Delta\omega_{\text{2D}}/\Delta\omega_{\text{G}}$ determined from eight different samples is on average 2.1. 
        The inset shows the absolute frequencies $\omega_{\text{2D}}$ and $\omega_{\text{G}}$ as reference.
    (c) 
        We observe the rupturing of the flake as a jump in $\omega_{\text{2D}}$ when the potential $V_a$ is swept.
        The frequency $\omega_{\text{2D}}$ typically jumps to a value above the one initially measured. 
    (d) The maximum shift $\Delta\omega_{\text{2D}}^{\text{max}}$ at which the graphene flake ruptures shows no correlation with the width of the flake.
        A false colored SEM image (similar to Figure 1b) of a ruptured graphene sheet (see inset) confirms the tearing at the edges.
    Each different sample has a unique color and symbol in panels (b) and (d).
  }
  \label{fig2}
\end{figure*}%
\begin{figure*}[!thp]
  \includegraphics[width=160mm]{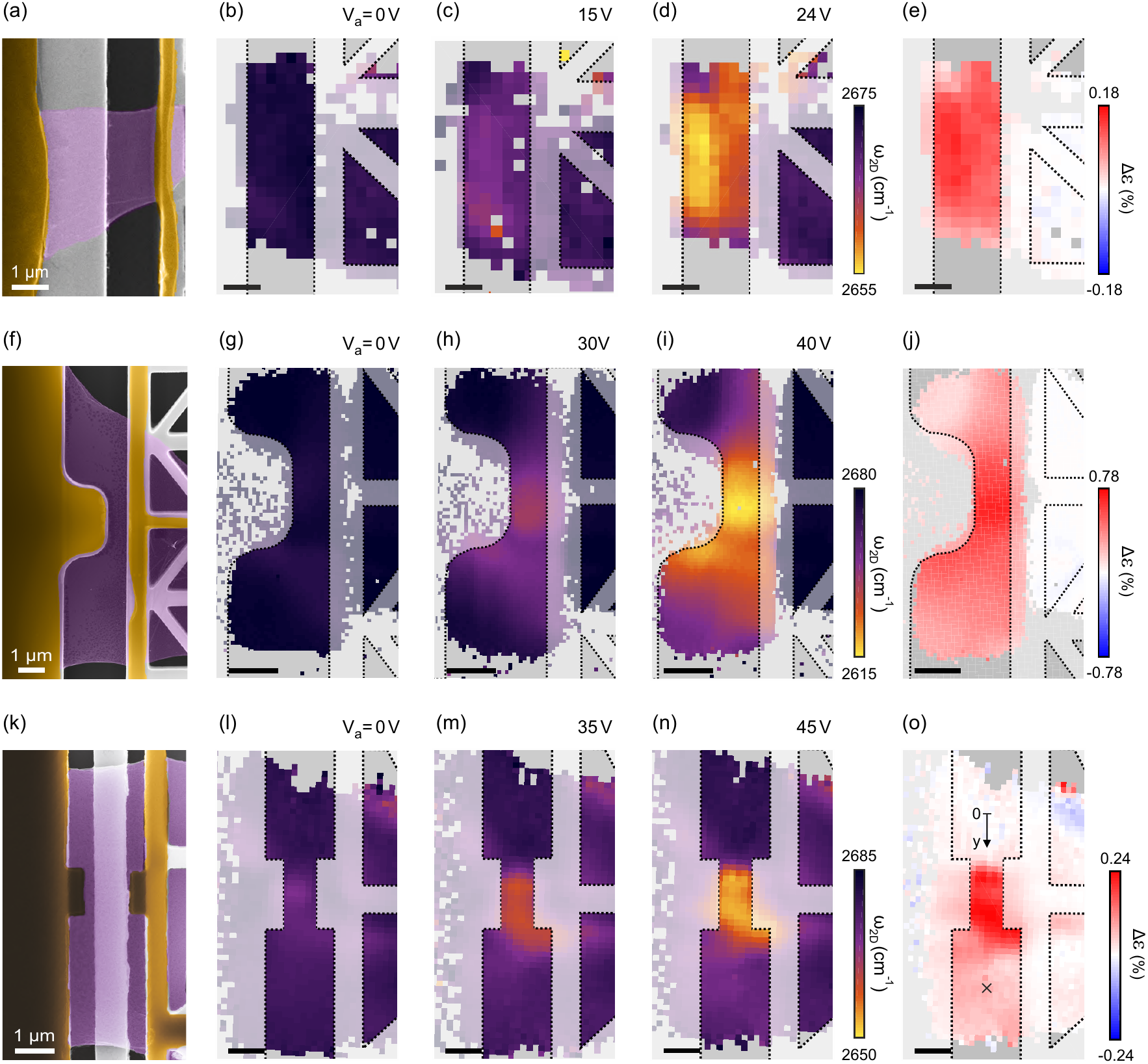}
  \vspace{-2mm}
  \caption{
    (a) 
            False color SEM image (similar to Figure 1b) of a typically measured device without noses.
    (b)-(d) Spatially resolved Raman maps of $\omega_{\text{2D}}$ for increasing $V_a$: (b) $V_a = 0$ V, (c) $V_a = 15$ V, and (d) $V_a = 24$ V.
            Note that these are taken on a different sample than the one shown in panel (a).
    (e)     The spatially resolved relative strain $\Delta \epsilon$ at $V_a = 24$ V.
 %
    (f)-(j) Similar as panels (a)-(e) for the CD design with a silicon nose.
            Please note the different $V_a$ values.
    (k)-(o) Similar as panels (a)-(e) for the CD design where the PMMA-clamping contains two noses.
    The CD design with the noses generate a strain hotspot at the position where the suspended length of the graphene is smallest.
    All black scale bars correspond to 1~\textmu m.
  }
  \label{fig3}
\end{figure*}

To monitor strain and to map spatially resolved strain fields, we use scanning confocal Raman microscopy (see Methods and Supplementary Figure~4).
Typical Raman spectra featuring the prominent Raman G and 2D peaks are shown in Figure~\ref{fig1}e and f, respectively.
We usually do not observe a D peak on the suspended part of the investigated graphene sheets.
The force $F$ applied to the suspended graphene flake results in a red shift of the G and 2D peak with increasing $|V_a|$ due to strain.
As the CD design allows the application of purely uniaxial strain, we observe a splitting of the G peak for high strain values (arrows in Figure~\ref{fig1}e), in good agreement with
Mohiuddin et al. \cite{mohiuddin2009}, which proves the strength of the CD actuators.
The splitting allows us to extract a Poisson ratio of $\nu = 0.11$ (Supplementary Figure~5), which is in good agreement with the one reported in the literature for graphite and suspended graphene \cite{mohiuddin2009}.

In the following, we focus on the low strain regime where both the G and 2D peak can be fitted by a single Lorentzian with center frequencies $\omega_{\text{G}}$ and
$\omega_{\text{2D}}$, respectively.
When sweeping $V_a$ back and forth many times (Figure~\ref{fig2}a and Supplementary Figure~6~and~7), we observe no hysteresis and hence no noticeable slipping of the clamped graphene.
The actuator thus induces strain in a controllable and reproducible fashion.
The center frequencies as a function of $V_a$ show a clear parabolic dependence (Figure~\ref{fig2}a) reflecting the linear stress-strain relation
\begin{equation}
\Delta \omega_{\text{2D}} = \frac{\partial \omega_{\text{2D}}}{\partial \epsilon} \Delta \epsilon = \frac{\partial \omega_{\text{2D}}}{\partial \epsilon} \frac{F}{k'L} = \frac{\partial
\omega_{\text{2D}}}{\partial \epsilon} \frac{\eta V_a^2}{k'L},
\end{equation}
where $\Delta \omega_{\text{2D}}$ is the shift in center frequency $\omega_{\text{2D}}$ ($\Delta \omega_{\text{G}}$ is analogue), $\partial \omega_{\text{2D}}/\partial \epsilon = -83$
${\rm cm^{-1}/\%}$ describes the relative $\omega_{\text{2D}}$ shift per unit strain $\epsilon$ (Supplementary Table~1), $\Delta \epsilon$ is the induced strain, and $k'$ a combined
spring constant, which takes into account the graphene flake suspended over the length $L$, the clamping and the actuator (Supplementary Discussion~1 and Supplementary Figures~8-10).

The shift in center frequencies are determined by the Gr\"uneisen parameters of both phonon modes \cite{mohiuddin2009}.
These Gr\"uneisen parameters quantify the anharmonicity of the lattice with respect to the direction of applied strain.
Consequently, the Gr\"uneisen parameters \cite{lee2012} are (slightly) dependent on the crystallographic orientation, which is seen in Figure~\ref{fig2}b as a variation in slopes
between $\Delta \omega_{\text{2D}}$ and $\Delta \omega_{\text{G}}$ for eight different devices.
The slope equals the ratio of the Gr\"uneisen parameters \cite{mohiuddin2009,lee2012} and we observe values ranging from 1.9 to 2.7 (see labels in Figure~\ref{fig2}b).
This agrees with the values between 1.9 and 3 reported in the literature \cite{lee2012}.

The maximum relative shift of $\omega_{\text{2D}}$ before mechanical failure is around 30 $\mathrm{cm^{-1}}$, which corresponds to at most $0.4\,\%$ of strain (inset
Figure~\ref{fig2}b).
Figure~\ref{fig2}c shows a typical example of a mechanical failure.
This particular graphene sheet teared while approaching $V_a = -24$ V.
This releases all strain, which results in a sudden hardening of $\omega_{\text{2D}}$, which jumps to values above the one at $V_a = 0$ V, due to the release of the existing pre-strain
introduced during the fabrication process.
SEM-pictures of broken devices (see e.g. the inset of Figure~\ref{fig2}d) confirm that the suspended graphene flake ruptured.
This implies that the PMMA-based clamping technique does not limit the maximum amount of strain that our samples can withstand.
The actuators can induce forces strong enough to rip the suspended graphene apart.
The surprisingly low rupturing point of the graphene flake (Figure~\ref{fig2}d) seems to be in contradiction with literature \cite{perez2014a,downs2016}.

To resolve strain patterns and to see where the graphene sheet ruptures, we use spatially resolved Raman spectroscopy.
The graphene sheet in Figure~\ref{fig3}a shows a uniform red shift of $\omega_{\text{2D}}$ with increasing $V_a$, due to strain (Figure~\ref{fig3}b-d).
The relative shift of $\omega_{\text{2D}}$ (Figure~\ref{fig3}e) illustrates the presence of strain at the edges.
Defects at the edges likely act as starting points for cracks propagating through the graphene flake when the overall strain is still low ($\sim 0.3\mathrm{\%}$) causing the mechanical
failure.

To reduce the strain at the edges and to get control over the strain field in the bulk of the suspended graphene flake, we modify the clamping geometry.
Here, we employed two different methods: (i) the CD actuator is designed with a nose and (ii) the pattern of the cross-linked PMMA includes two noses.
The noses locally reduce the distance between the fixed anchor and the suspended body of the CD actuator, which results in a higher strain in relation to the strain at the edge.
Typical results for methods (i) and (ii) are shown in Figure~\ref{fig3}f-j and Figure~\ref{fig3}k-o, respectively.
Both methods show strain hotspots located at positions where the distance between the fixed anchor and the suspended body is minimal.
The strain in the hotspots is up to 4 times larger than at the edges.
Note the presence of a strain hotspot at $V_a = 0$ V for method (ii).
We reached a maximum strain of 1.2\% in the hotspots, which largely exceeds the rupturing strain of $0.3\mathrm{\%}$ for the devices shown in Figure~\ref{fig2}d.
This not only highlights the crucial role of clamping in translating the applied force to strain, but also reinforces the conclusion that the edges are responsible for the low
rupturing point.

\begin{figure}[!thp]
  \includegraphics[width=85mm]{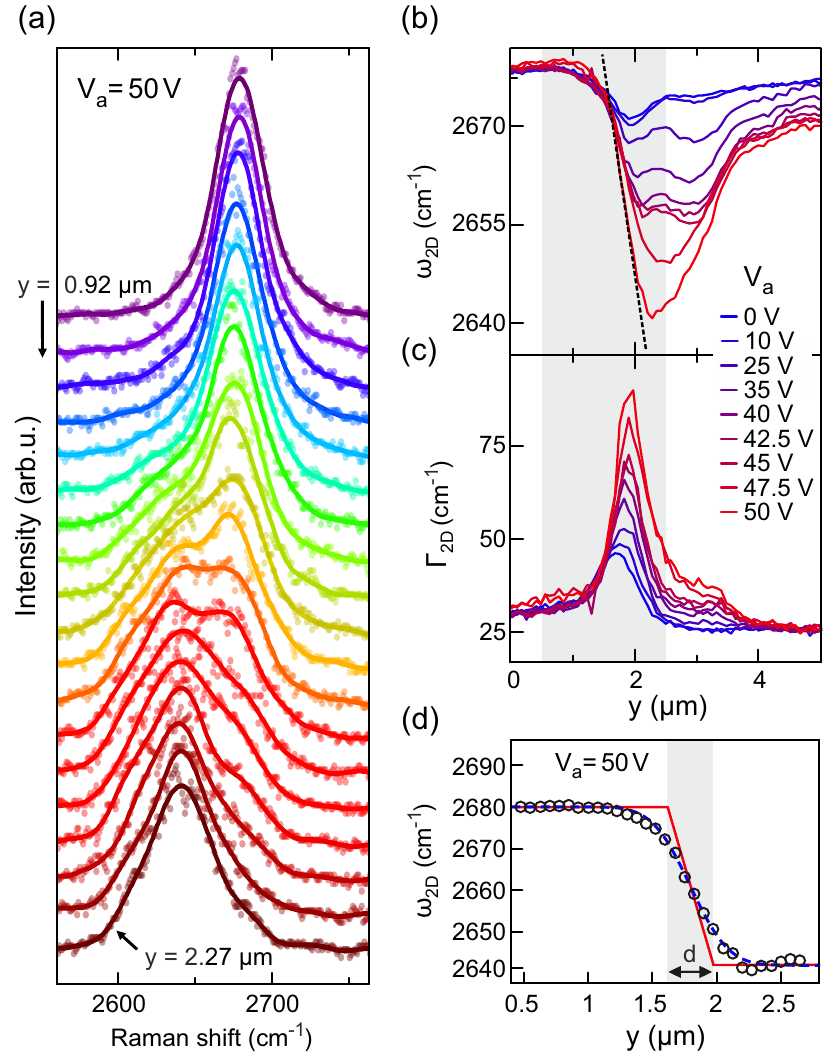}
  \vspace{-2mm}
  \caption{
    (a) Raman spectrum around the 2D peak along the y-direction highlighted in Figure~\ref{fig3}o (at $V_a = 50$ V).
    (b) $\omega_{\text{2D}}$ along the y-direction as shown in Figure~\ref{fig3}o (see cross for end point) for different applied potentials $V_a$.
        The black dashed line indicates the maximum strain gradient.
    (c) $\Gamma_{\text{2D}}$ shows a strong inhomogeneous broadening at the positions where the strain gradient in panel (b) is maximum.
    (d) To quantify the strain gradient in the grey shaded area in panels (b) and (c), we convolute a piecewise linear function with width $d$ (arrow) with the laser spot.
        The fit result gives us a width $d$ of 325 nm, which corresponds to a gradient $\partial\omega_{\text{2D}}/\partial y=117\,\mathrm{cm^{-1}}$/\textmu m or $\partial
        \epsilon/\partial y = 1.4$ \%/\textmu m at $V_a = 50$ V.
  }
  \label{fig4}
\end{figure}%
\begin{figure*}[!thp]
  \includegraphics[width=178mm]{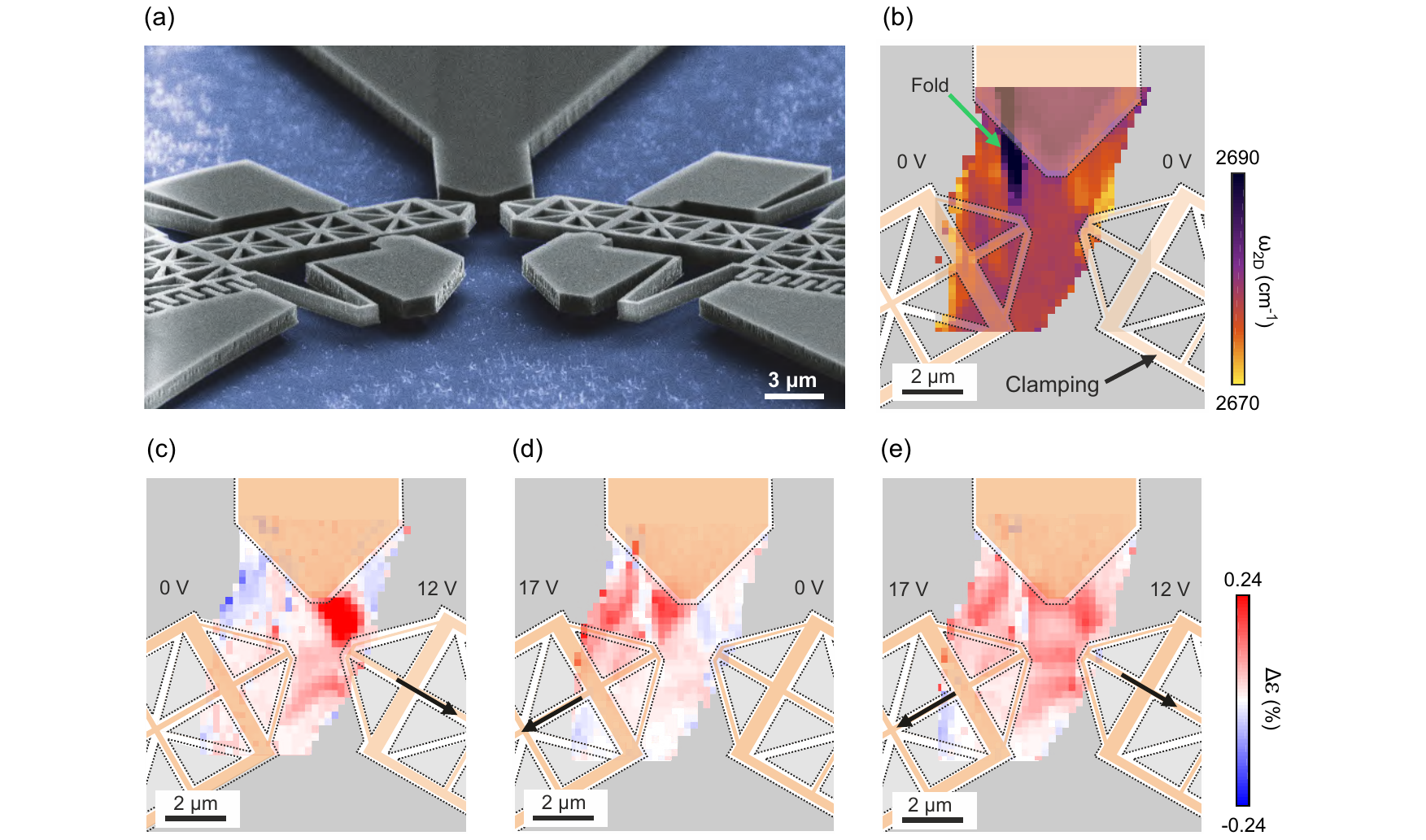}
  \vspace{-2mm}
  \caption{
    (a) False color SEM image of our fabricated MEMS device with two CD actuators (grey).
    (b) Spatially resolved Raman map of $\omega_{\text{2D}}$ for $V_a = 0$ V on both actuators.
    (c)-(e) The spatially resolved relative strain $\Delta \epsilon$ for increasing $V_a$: (c) $V_a = 0$ V on the left actuator and $V_a = 12$ V on the right actuator, (d) $V_a = 17$ V
    on the left actuator and $V_a = 0$ V on the right actuator, and (e) $V_a = 17$ V on the left actuator and $V_a = 12$ V on the right actuator.
    Note the blue shift of $\omega_{\text{2D}}$ if 0 V is applied (panels (c) and (d)) to one of the CD actuators, which corresponds to a relative compression of the graphene flake. 
    The cross-linked PMMA areas are shaded orange in panels (b)-(e).
  }
  \label{fig5}
  \vspace{-5mm}
\end{figure*}

The control over strain fields also allows to design and control strain \emph{gradients}.
Figure~\ref{fig4}a shows that the 2D peak frequency decreases when scanning over the strain gradient in the y-direction of highest strain (see line in Figure~\ref{fig3}o).
This frequency shift is shown in Figure~\ref{fig4}b as a function of position.
The highest strain gradients are on the left flank of the hotspot (Figure~\ref{fig4}b), which corresponds to the upper edge of the PMMA-jut in Figure~\ref{fig3}o.
In the region of highest strain gradient, there is also a clear broadening of the 2D peak, which cannot be associated with a splitting into two sub-peaks \cite{yoon2011}, since the
widths of the peaks in the regions of high and low strain are similar (Figure~\ref{fig4}c).
We attribute it instead to inhomogeneous broadening, due to the fact that, in the region of strong strain gradient, the laser spot sample areas with different strain conditions
\cite{neumann2015b,banszerus2017}.

Consequently, we have to deconvolve the extracted $\omega_{\text{2D}}$ (Figure~\ref{fig4}a) into a piecewise-linear function of width $d$ (Figure~\ref{fig4}d) and the laser spot, which
is modeled by a Gaussian function with a diameter of $505 \pm 10$ nm (Supplementary Figure~11) to determine the actual strain profile.
At $V_a=50$~V, we find $d = 325$~nm, which corresponds to a gradient of $\partial \omega_{\text{2D}}/\partial y = 117$ $\mathrm{cm^{-1}}$/\textmu m and thus to a strain gradient of
$\partial \epsilon/\partial y = 1.4$ \%/\textmu m.
Assuming the strain is applied in the armchair direction \cite{pereira2009b}, a strain gradient of 1.4 \%/\textmu m induces a pseudomagnetic field of
$B_{\text{ps}}=\hbar\beta/2ae\cdot\partial\epsilon/\partial y = 120\,\mathrm{mT}$, where $\beta=3.37$ is the logarithmic derivative of the nearest-neighbour hopping parameter and
$a=1.42\,\mathrm{\mathring{A}}$ is the nearest-neighbour distance.
This $B_{\text{ps}}$ is strong enough to bring state-of-the-art ultra-clean graphene devices \cite{novoselov2004,banszerus2015,dauber2015} into the quantum Hall regime, which is
considered a crucial step for the realization of valley-tronics \cite{ma2016}.

A very promising geometry for a uniform pseudomagnetic field in graphene requires triaxial strain fields \cite{verbiest2015}.
We realize such strain fields by integrating graphene on two 120$^{\circ}$ rotated CD actuators placed close to each other (Figure~\ref{fig5}a).
The $\omega_{\text{2D}}$ Raman maps in Figure~\ref{fig5}b-e show the independent control of the strain in graphene clamped onto the two different CD actuators and the fixed anchor.
Interestingly, a single CD actuator not only pulls the graphene sheet but also the other CD actuator, which results in a relative blue shift of the $\omega_{\text{2D}}$ frequency and
thus locally in compressive strain (Figure~\ref{fig5}c and d).
By applying approximately equal forces with both CD actuators, we obtain a triaxial strain field in the center region (Figure~\ref{fig5}e).

In conclusion our approach allows to induce well controlled and tunable strain fields in two-dimensional materials without any spurious effect due to capacitive coupling to the
suspended flake or interaction with the substrate.
Understanding and controlling strain in two-dimensional materials will not only advance fundamental knowledge of the unique electromechanical coupling of graphene and other 2D
materials, but will also enable ways for dynamically manipulating carriers for novel device applications.

{\bf Methods.}
\emph{Experimental methods and details.}
We fabricate the micro-actuators on a silicon on insulator (SOI) substrate following the process described in Ref.\cite{goldsche2014}. The substrate consists from bottom to top of a
500 \textmu m thick Si layer, a 1 \textmu m thick SiO$_{\mathrm{2}}$ layer, and a 2 \textmu m chemical vapor deposited crystalline, highly p-doped silicon layer. The doping of the top
layer is $10^{19}$ cm$^{-3}$, making our devices low temperature compatible \cite{verbiest2016}. The process flow is depicted in Supplementary Figure~1. Using standard electron beam
lithography techniques and reactive ion etching with C$_{\mathrm{4}}$F$_{\mathrm{8}}$  and SF$_{\mathrm{6}}$, we pattern the CD actuators, shown in Figure~\ref{fig1}b~and~c. The
graphene crystals are prepared by mechanical exfoliation from bulk graphite on top of PMMA. Consequently, we transfer the graphene-PMMA stack onto the SOI substrate
\cite{mayorov2011,dean2010} with an accuracy of $\sim 1$~\textmu m using micro-manipulators and an optical microscope. Afterwards, the graphene is fixed to the actuator by
cross-linking the top PMMA layer at certain locations (see Figure~\ref{fig1}b). After dissolving the not cross-linked PMMA with acetone, we suspend the micro-actuator by etching the
SiO$_{\mathrm{2}}$ underneath with 10\% hydrofluoric acid solution. Finally, a critical-point drying step is used to prevent the CD actuator from collapsing due to capillary forces.

\emph{Raman measurements.}
We use a partly home-built low temperature ($\sim 4$~K) Raman setup that contains a laser with a wavelength of $\lambda=532\,\mathrm{nm}$, which is focused onto the sample by an
$100\times$ objective with a numerical aperture of 0.82. We use a laser intensity of 1~mW and the laser spot diameter is approximately $505 \pm 10\,\mathrm{nm}$ (Supplementary
Figure~11). All measurements in this manuscript were performed with linearly polarized light. The reflected and scattered light is detected via a single-mode optical fiber and a
spectrometer with a grating of 1200~lines/mm. For performing spatially resolved Raman spectroscopy, our sample is mounted on top of a dc $xy$-piezo stage. The low temperature minimizes
the drift during the measurements. We always measure first the Raman spectrum from 250 $\mathrm{cm^{-1}}$ up to 1750 $\mathrm{cm^{-1}}$ to simultaneously obtain the silicon peak at 521
$\mathrm{cm^{-1}}$ and the G peak around 1584 $\mathrm{cm^{-1}}$. We use in-house developed machine learning algorithm to align further Raman maps, which include the 2D peak, with the
CD actuator (Supplementary Figure~4).

\section*{Associated content}
\subsection*{Supporting Information}
\noindent
Details on the fabrication, the capacitive coupling, and the behavior of CD actuators as well as details on the Raman pre-characterization of integrated graphene flakes, the estimation
of the Poisson ratio, the measurement reproducibility, the combined spring constant $k'$, the determination of the laser spot size, and the reported Raman peak shifts due to strain,
are available free of charge via the Internet at \url{http://pucs.acs.org}.

\section*{Author information}
\subsection*{Corresponding author}
\noindent
E-mail: stampfer@physik.rwth-aachen.de


\subsection*{Notes}
\noindent
The authors declare no competing financial interests.

\section*{Acknowledgments}
\noindent
The authors would like to thank S. Trellenkamp, B. Hermanns, M. Muoth, F. Hassler, F. Haupt for valuable discussions.
Support by the ERC (GA-Nr. 280140), the Helmholtz Nanoelectronic
Facility (HNF) \cite{hnf2017} at the Forschungszentrum J\"ulich, and the Deutsche Forschungsgemeinschaft (DFG) (SPP-1459) are gratefully acknowledged.
S.R. acknowledges funding by the National Research Fund (FNR) Luxembourg (project RAMGRASEA).
G.V. acknowledges funding by the Excellence Initiative of the German federal and state governments.

%


\providecommand{\latin}[1]{#1}
\makeatletter
\providecommand{\doi}
{\begingroup\let\do\@makeother\dospecials
	\catcode`\{=1 \catcode`\}=2 \doi@aux}
\providecommand{\doi@aux}[1]{\endgroup\texttt{#1}}
\makeatother
\providecommand*\mcitethebibliography{\thebibliography}
\csname @ifundefined\endcsname{endmcitethebibliography}
{\let\endmcitethebibliography\endthebibliography}{}


\newpage
\clearpage

\onecolumngrid
\setcounter{figure}{0}
\renewcommand{\figurename}{Supplementary Figure}
\renewcommand{\tablename}{Supplementary Table}
\renewcommand{\thetable}{\arabic{table}}

{\LARGE {\bf Supporting information:\\Tailoring mechanically-tunable strain fields in graphene}}
\ \\ \ \\
Supplementary Discussion 1
\ \\ \ \\
The effective spring constant $k'$ (see Eq.~1 in the main manuscript) contains the spring constant $k_{\text{a}}$ of the actuator, the spring constant $k_{\text{P}}$ of the (cross-linked) PMMA clamping, and the spring constant $k$ of the graphene flake (see Supplementary Figure~8a).
It describes how much of the force $F = \eta V_a^2$ exerted by the actuator is transduced into the graphene sheet.
Considering the spring configuration depicted in Supplementary Figure~8a, we find:

\vspace{-3mm}
\begin{equation}
k' = k \left( 1 + k_{\text{a}} \left[ \frac{1}{k} + \frac{1}{k_{\text{P}}} \right] \right).
\label{seq1}
\end{equation}

It is instructive to see that Eq.~\ref{seq1} reduces to $k' \approx k$ if both following conditions are met: (i) ideal clamping ($k/k_{\text{P}} <\!\!< 1$) and (ii) soft actuator ($k_{\text{a}}/k <\!\!< 1$), which means that all the force exerted by the actuator is directly transduced to the graphene sheet.

We use $k = Y W/L$ to estimate the spring constant of a graphene flake of length $L = 2$ \textmu m and width $W$.
We assume a two-dimensional Young's modulus $Y$ of 362 N/m, as measured by Lee~\textit{et~al.} \cite{lee2008}.

The spring constant $k_{\text{a}}$ of the actuator is estimated from COMSOL simulations \cite{comsol} (see for example Supplementary Figure~3).
For all CD devices used in this work, we found that $k_{\text{a}}$ is below 15 N/m.
Considering the widths of the graphene flakes used in this work (Figure~2d in the main manuscript and Supplementary Figure~9), we find that $k$ is always larger than 500 N/m.
Therefore, we conclude that condition (ii) is always fulfilled.

To estimate the ratio $k'/k$, we use Eq. 1 in the main text to extract $k'$ from fitting $\Delta\omega_{\text{2D}}(V_a)$ to our experimental data (see Supplementary Figure~9).
Here we use $\partial \omega_{\text{2D}}/\partial \epsilon = -83\,\text{cm}^{\text{-1}}/\text{\%}$.
The ratio $k'/k$, listed in Supplementary Figure~9g, is for all devices larger than 1.8.
We point out that this is unlikely explained by a different Young's modulus for the graphene, as it would require a value that is almost twice the literature value (362 N/m).
Therefore, we conclude that condition (i) is not met.

Based on this, we can estimate the spring constant of our clamping by inverting Eq.~\ref{seq1} into an equation for $k_{\text{P}}$:

\vspace{-3mm}
\begin{equation}
k_{\text{P}} = \frac{k_{\text{a}}}{k'/k - k_{\text{a}}/k - 1}.
\label{seq2}
\end{equation}

When considering the clamping geometry (Supplementary Figure 8c,d), we can define a lower and an upper bound for $k_{\text{P}}$ from Euler beam theory.
The cross-linked PMMA beams covering the width of the graphene flake give a lower bound for $k_{\text{P}}$:

\vspace{-5mm}
\begin{equation}
k_{\text P}^{\text{low}} = \frac{16 E b^3 h / 3}{(W + W_{\text{offset}})^3},
\label{seq3}
\end{equation}

where $E = 4$ GPa is the Young's modulus of cross-linked PMMA \cite{teh2003,torres2010}, $W_{\text{offset}}$ = 2 \textmu m, and $b$ is the width of the cross-linked PMMA beam with thickness $h$ (see Supplementary Figure~8c,d).
We included the $W_{\text{offset}}$ to account for the effect that a small part of the PMMA beam on the silicon is deformed.
We estimated the value of $W_{\text{offset}} \approx 2$ \textmu m from the scanning electron microscope image in Supplementary Figure~8c,d by the length of the white lines.
The beams along the length of the graphene flake give an upper bound for $k_{\text{P}}$:

\vspace{-3mm}
\begin{equation}
k_{\text{P}}^{\text{high}} = \frac{E b h}{L_{\text{overlap}}},
\label{seq4}
\end{equation}

where $L_{\text{overlap}}$ is the length of the beam on the graphene (see Supplementary Figure 8c).
In this simple picture, we assume a perfect adhesion of the cross-linked PMMA to the silicon and zero adhesion of the graphene to the silicon.
In addition, we neglect the fact that the graphene is clamped by multiple beams covering the width of the graphene.
Nevertheless, we find that the experimentally extracted $k_{\text{P}}$ is always between the theoretical lower and upper bound (see Supplementary Figure~10).
Interestingly, sample number 1 (Supplementary Figure 8d), which is closest to the lower bound, did not have a clamping beam along the length of the graphene (see Supplementary Figure 8c for an example).
All the other samples had a beam along the length of the graphene and are therefore closer to the upper bound.

We conclude that the clamping is elastic and acts as a spring in series to $k$, which are both parallel to the spring of the actuator.
The experimentally determined spring constant of the clamping is well within theoretical bounds given by Euler beam theory and the clamping geometry.

Finally, we point out that $\partial \omega_{\text{2D}}/\partial \epsilon = -83\,\text{cm}^{\text{-1}}/\text{\%}$ will also slightly depend on the crystallographic orientation of the graphene sheet with respect to the pulling direction of the comb-drive. However, this effect is small and therefore cannot explain the observed $k'/k$ ratio.

\newpage

\begin{figure}[th]
  \includegraphics[width=99.2mm]{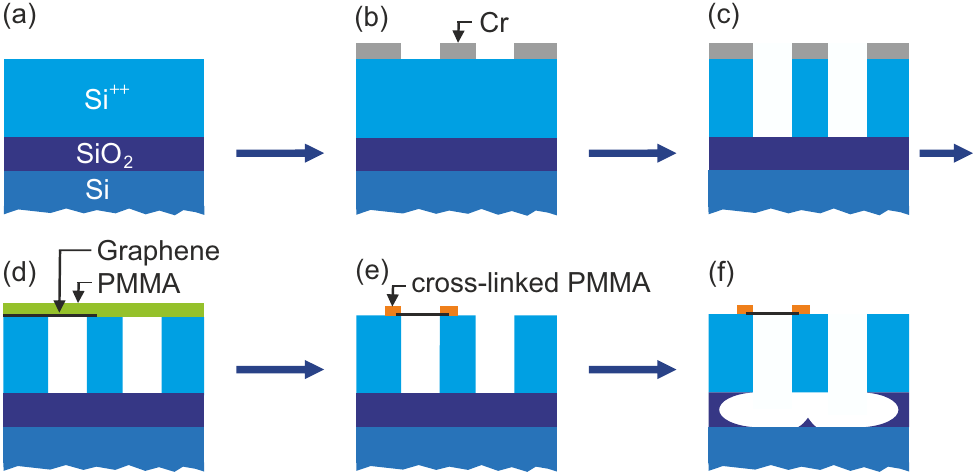}
  \caption{
    Fabrication of comb-drive (CD) actuators with integrated suspended graphene.
    (a) The CD actuators are fabricated on a silicon on insulator (SOI) substrate that consists from bottom to top of a 500 \textmu m thick Si layer, a 1 \textmu m thick
        SiO$_{\mathrm{2}}$ layer, and a 2 \textmu m chemical vapor deposited crystalline, highly p-doped silicon layer.
    (b) By using standard electron beam lithography techniques, we deposit a Cr hard mask to outline the CD actuator.
    (c) We pattern the CD actuators via reactive ion etching with C$_{\mathrm{4}}$F$_{\mathrm{8}}$  and SF$_{\mathrm{6}}$.
    (d) The prepared graphene-PMMA stack is transferred onto the SOI substrate with an accuracy of $\sim 1$~\textmu m.
    (e) A second electron beam lithography step is used to cross-link the top PMMA layer to clamp the graphene on the CD actuator.
        The not cross-linked PMMA is dissolved with acetone.
    (f) To suspend the CD actuator with the integrated suspended graphene, we etch away the SiO$_{\mathrm{2}}$ layer with a 10\% hydrofluoric acid solution and a critical point
        drying step is used to prevent the CD actuator from collapsing.
  }
  \label{Sub_fig1}
\end{figure}

\ \\ \ \\ \ \\ \ \\ \ \\ \ \\ \ \\ \ \\
\ \\ \ \\ \ \\ \ \\ \ \\ \ \\ \ \\ \ \\
\ \\ \ \\ \ \\ \ \\ \ \\ \ \\ \ \\ \ \\

\newpage

\begin{figure}[th]
\includegraphics[width=178mm]{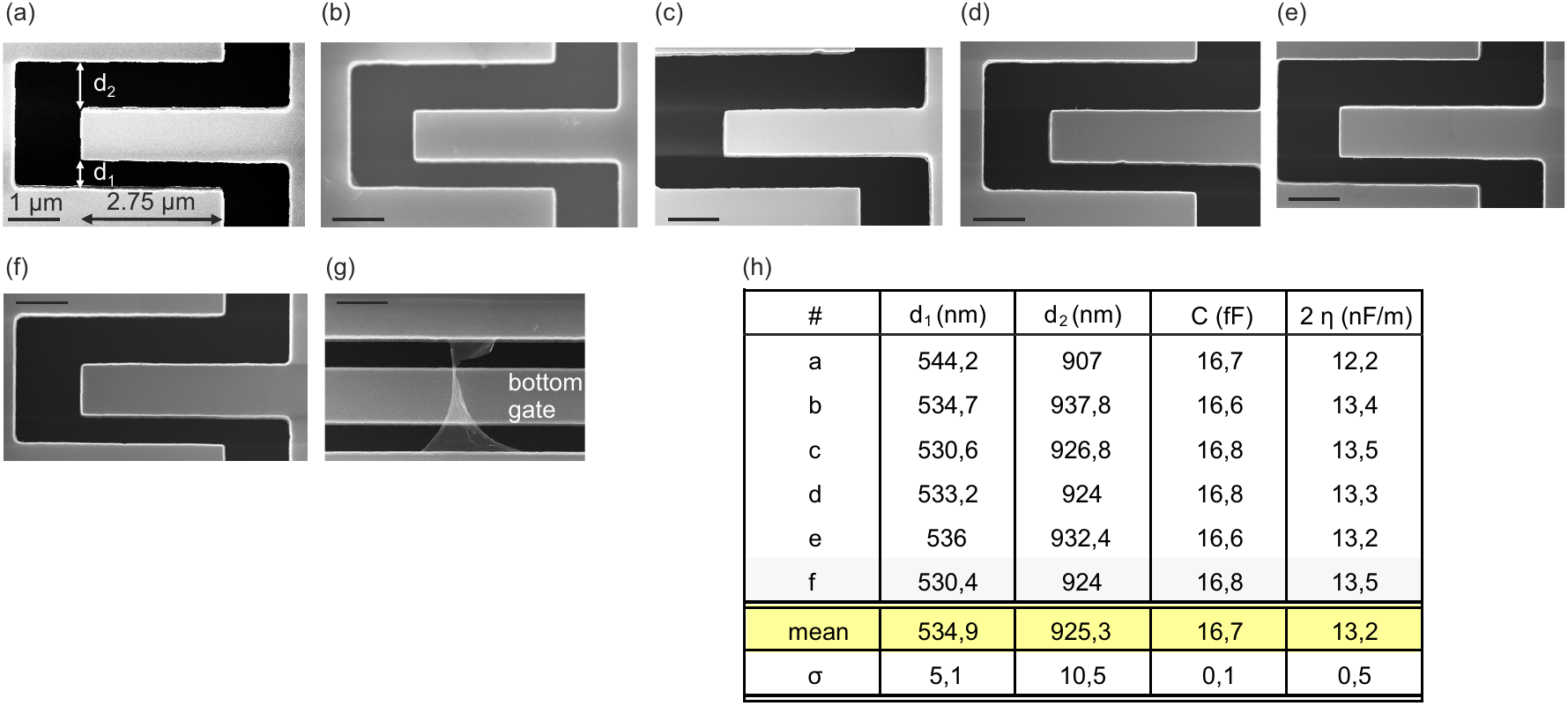}
  \caption{
    Estimation of the capacitive coupling between the interdigitated fingers.
    We design the distance between the fingers of the CD actuators to be $d_1 = 500$ nm and $d_2 = 900$ nm (see white arrows and labels in panel (a)).
    To estimate the capacitive coupling between the fingers, we measure these distances for a number of fabricated devices.
    Panels (a) to (e) show the distances of devices without any graphene and panel (f) shows the distances of a device with integrated graphene as indicated in panel (g).
    Note that the graphene in panel (g) is suspended over an electrostatic bottom gate that is placed $\sim 300$ nm underneath the suspended graphene sheet.
    The scale bars are all 1 \textmu m long.
    All measured distances are listed in panel (h).
    We find that all distances are slightly higher than designed.
    Note that the distances extracted from the device with integrated graphene is no different from the ones measured on devices without graphene.
    Using the finger overlap area $A$ as well as the number of fingers $N = 118$, we find a capacitance $C = \epsilon_0 A N (d_1^{-1} + d_2^{-1})$ of around $16.7\pm 0.1$ fF between the fingers.
    The capacitive coupling $2\eta = \epsilon_0 A N (-d_1^{-2} + d_2^{-2})$ relevant for the electrostatic force (see main manuscript) is therefore $13.2 \pm 0.5$ nF/m (or $2\eta = 13.2 \pm 0.5 \mathrm{nN/V^2}$).
  }
  \label{Sup_fig7v1}
\end{figure}

\newpage

\begin{figure}[th]
  \includegraphics[width=94.3mm]{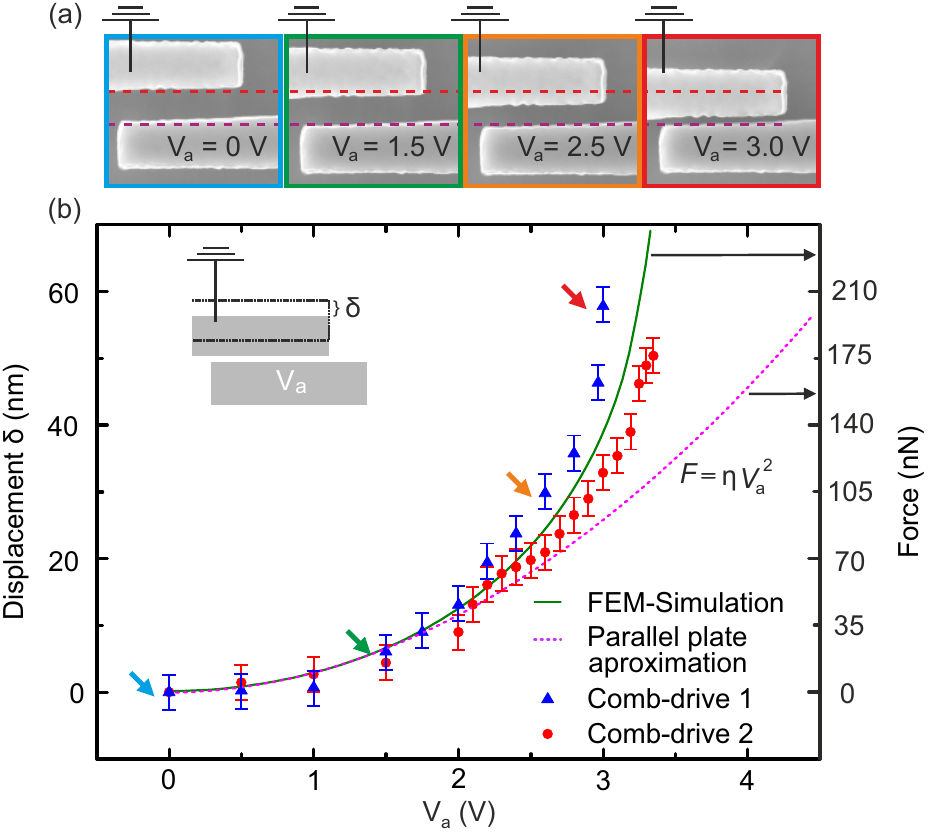}
  \caption{
    Scanning electron microscope (SEM) characterization of empty CD actuators.
    (a) SEM images of the separation between the interdigitated fingers of an empty CD actuator (without a graphene flake) as a function of applied potential $V_a$.
    (b) The displacement $\delta$ (see inset) of two different CD actuators extracted from SEM images for increasing $V_a$.
        The arrows indicate the corresponding SEM image in panel (a).
        The displacement $\delta$ shows a nearly quadratic dependence on applied potential $V_a$, which is consistent with the electrostatic force.
        To estimate this force, we performed COMSOL simulations with the fabricated device geometry \cite{comsol}.
        The extracted spring constant of the CD actuator from the simulations is $\approx$ 3.5 N/m.
        The green line depicts the simulation result and indicates a maximum force of $\approx$ 200 nN at a displacement of 60 nm.
        The pink line shows that the force obtained using the parallel plate approximation for the capacitance is in good agreement with the simulation for displacements up to 10 nm.
        Please note that $2\eta = 8.7 \pm 0.5$ nF/m for these devices, as they had a slightly different design than the devices shown in Supplementary Figure~2.
  }
  \label{Sub_fig2v}
\end{figure}

\newpage

\begin{figure}[th]
  \includegraphics[width=131.5mm]{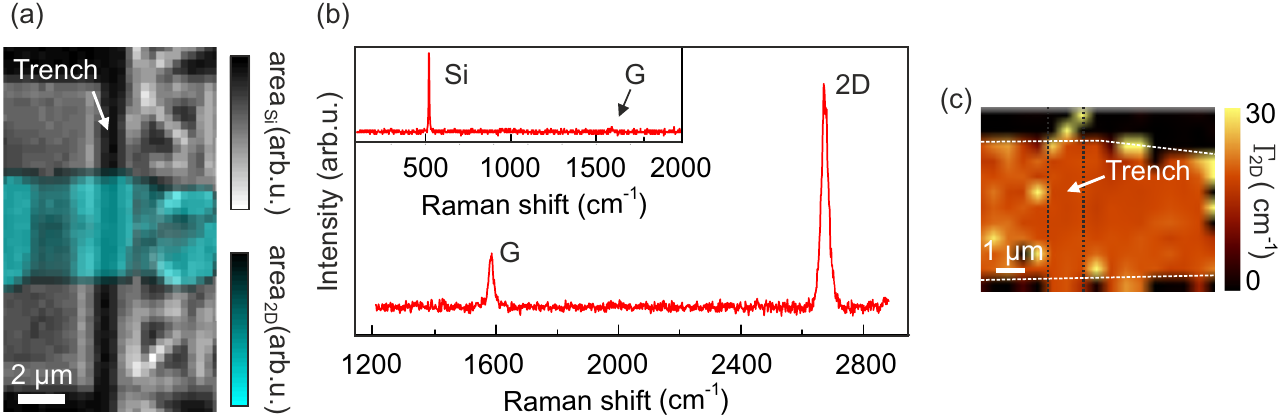}
  \caption{
    Raman pre-characterization.
	(a) Raman map of the silicon peak area at 521 $\mathrm{cm^{-1}}$ superimposed with a Raman map of the 2D peak area, which has been obtained as follows.
        In a first Raman map, we measure from 250 $\mathrm{cm^{-1}}$  to 1750 $\mathrm{cm^{-1}}$, which gives us both the silicon peak and the G peak of the graphene.
        In a second Raman map, we measure both the G and 2D peak.
        By aligning both measured Raman maps of the G peak, we can superimpose the 2D peak with the silicon peak.
	(b) Typical Raman spectrum of a suspended graphene flake, which was transferred on one of our CD actuators.
        Apart from the visible G and 2D peak, it is important to note the absence of a D peak (expected around 1345 $\mathrm{cm^{-1}}$).
        The inset shows the Raman spectrum around the silicon peak at 521 $\mathrm{cm^{-1}}$ that is obtained on a CD actuator.
    (c) The Raman map of the width $\Gamma_{\text 2D}$ of the 2D peak obtained on a typical sample.
  }
  \label{Sup_fig3}
\end{figure}

\newpage

\begin{figure}[th]
  \includegraphics[width=172.3mm]{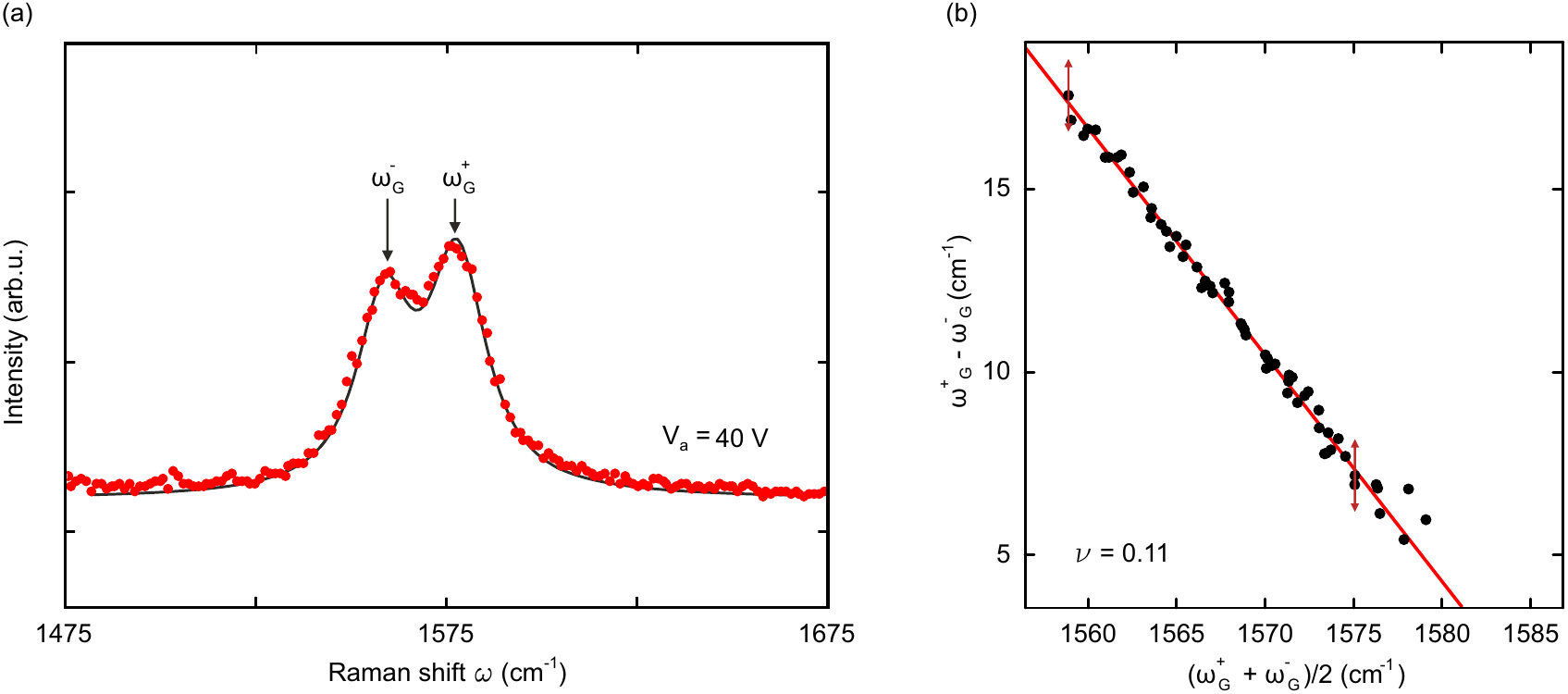}
  \caption{
    Estimation of the Poisson ratio.
	(a) The Raman spectrum of the G peak for $V_a = 40$ V shows a splitting into two peaks due to uniaxial strain as depicted in Figure~1e of the main manuscript.
        These peaks are the so-called $\omega^{-}_{\text G}$ and $\omega^{+}_{\text G}$ peak at lower and higher Raman shift, respectively \cite{mohiuddin2009}.
    (b) The frequency difference $\omega^{+}_{\text G} - \omega^{-}_{\text G}$ as a function of their average $(\omega^{+}_{\text G} + \omega^{-}_{\text G})/2$ allows us to extract the Poisson ratio of the suspended graphene sheet according to the following equation \cite{mohiuddin2009}:
    \newline
    \newline
    $\displaystyle \omega^{+}_{\text G} - \omega^{-}_{\text G} = -\frac{0.99}{1.99} \frac{1+\nu}{1-\nu} \left( \frac{\omega^{+}_{\text G} + \omega^{-}_{\text G}}{2} - \omega^{0}_{\text G} \right)$,
    \newline
    \newline
    where $\nu$ is the Poisson ratio and $\omega^{0}_{\text G}$ is the $\omega_G$ frequency at zero strain.
    By fitting our data between the vertical arrows with a linear function (see red line), we obtain a Poisson ratio of $\nu = 0.11$.
    The extracted Poisson ratio is in good agreement with the one reported in the literature for graphite and suspended graphene ($\nu = 0.13$) \cite{mohiuddin2009}.
  }
  \label{Sup_fig8v1}
\end{figure}

\newpage

\begin{figure}[!ht]
\includegraphics[width=158.5mm]{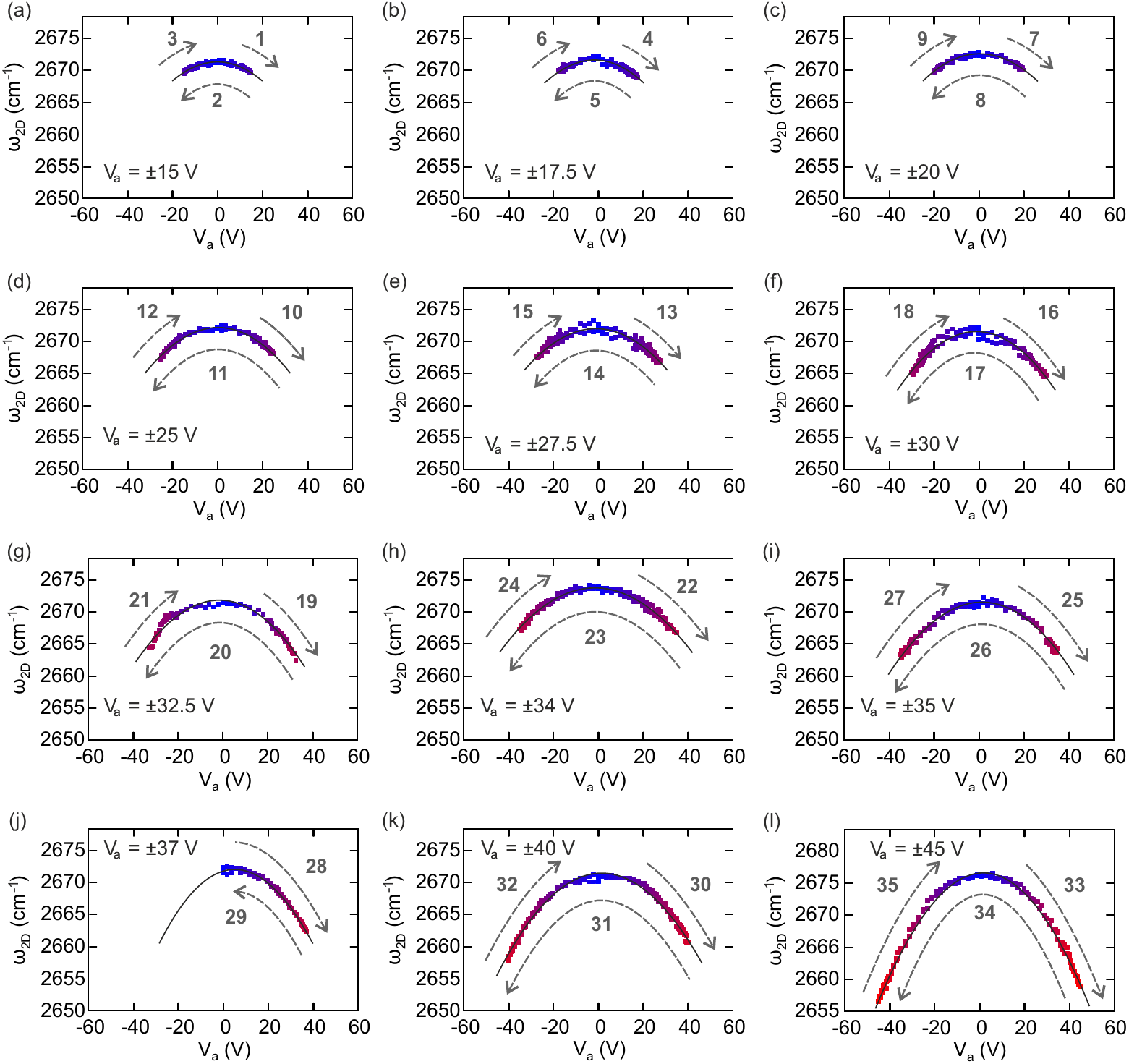}
  \caption{
    Measurement reproducibility.
    We show the extracted $\omega_{\text{2D}}$ frequency obtained on the same position of the graphene flake for different measurement cycles.
    The measurements are depicted in chronological order from panel (a) to (l).
    In each cycle, we sweep the potential $V_a$ back and forth (gray arrows) while monitoring the Raman spectra.
    After each cycle, we increase the maximum applied $|V_a|$ and start a new cycle.
    We observe no hysteresis and hence no noticeable slipping of the clamped graphene.
    Hence, we can combine all cycles into a single graph as was done in Figure~2 of the main manuscript.
    This allows us to obtain a unprecedent amount of data points on each sample.
    Note that the scale in panel (l) is different from those in (a)-(k).
    After measuring panel (k), we moved the sample into a different measurement system, which resulted in an overall shift of $\omega_{\text{2D}}$.
    Surprisingly, the tuning with $V_a$ did not change at all indicating the mechanical stability of our samples.
  }
  \label{Sup_new_sup_fig_b}
\end{figure}

\newpage

\begin{figure}[!ht]
\includegraphics[width=157.8mm]{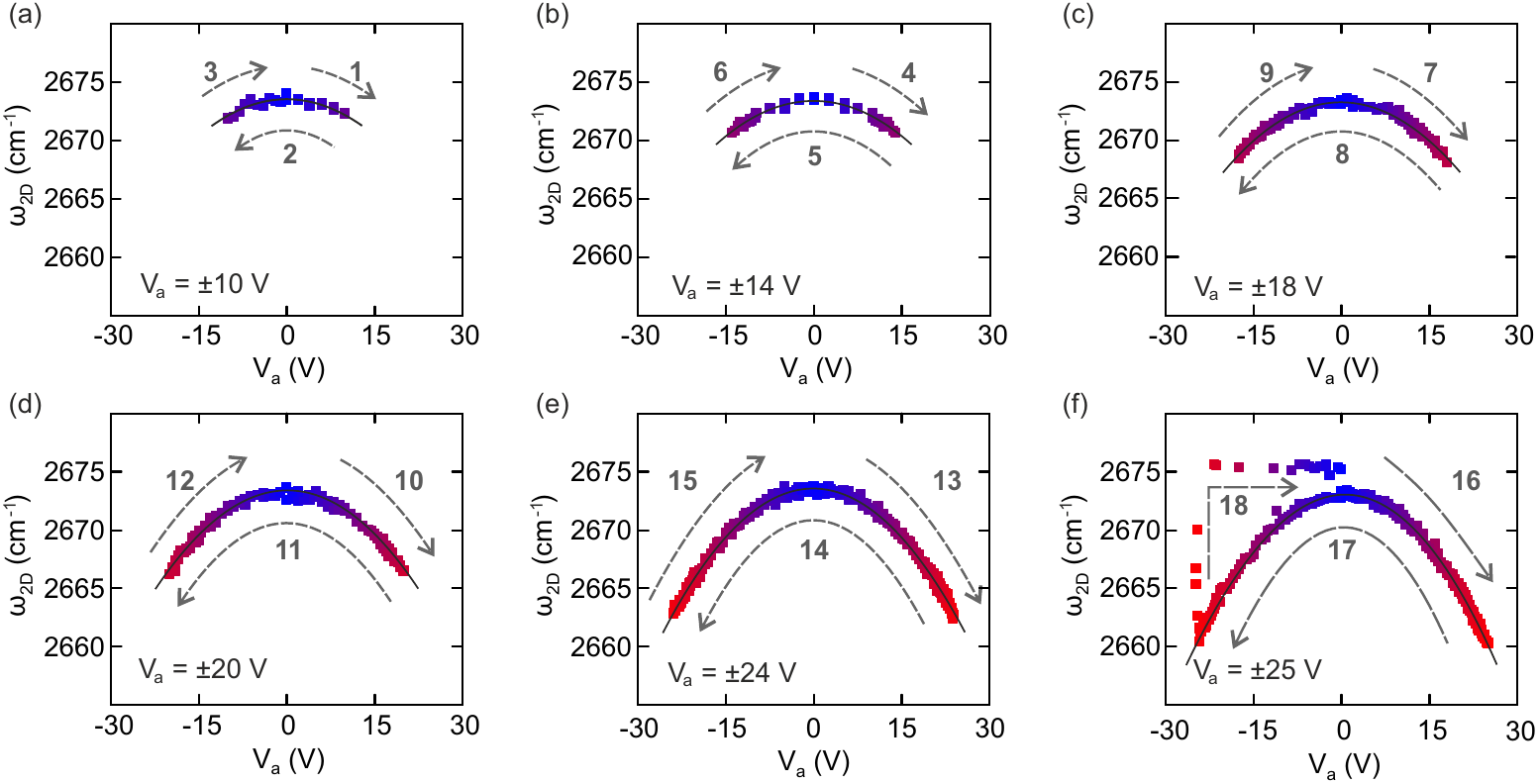}
  \caption{
    Measurement reproducibility.
    We show the extracted $\omega_{\text{2D}}$ frequency obtained on the same position of the graphene flake for different measurement cycles.
    The measurements are depicted in chronological order from panel (a) to (f).
    In each cycle, we sweep the potential $V_a$ back and forth (gray arrows) while monitoring the Raman spectra.
    After each cycle, we increase the maximum applied $|V_a|$ and start a new cycle.
    By increasing $|V_a|$ to 25 V, we observe the rupturing of the graphene flake (see Figure~2c in the main text).
  }
  \label{Sup_new_sup_figa}
\end{figure}

\newpage

\begin{figure}[!ht]
\includegraphics[width=93.2mm]{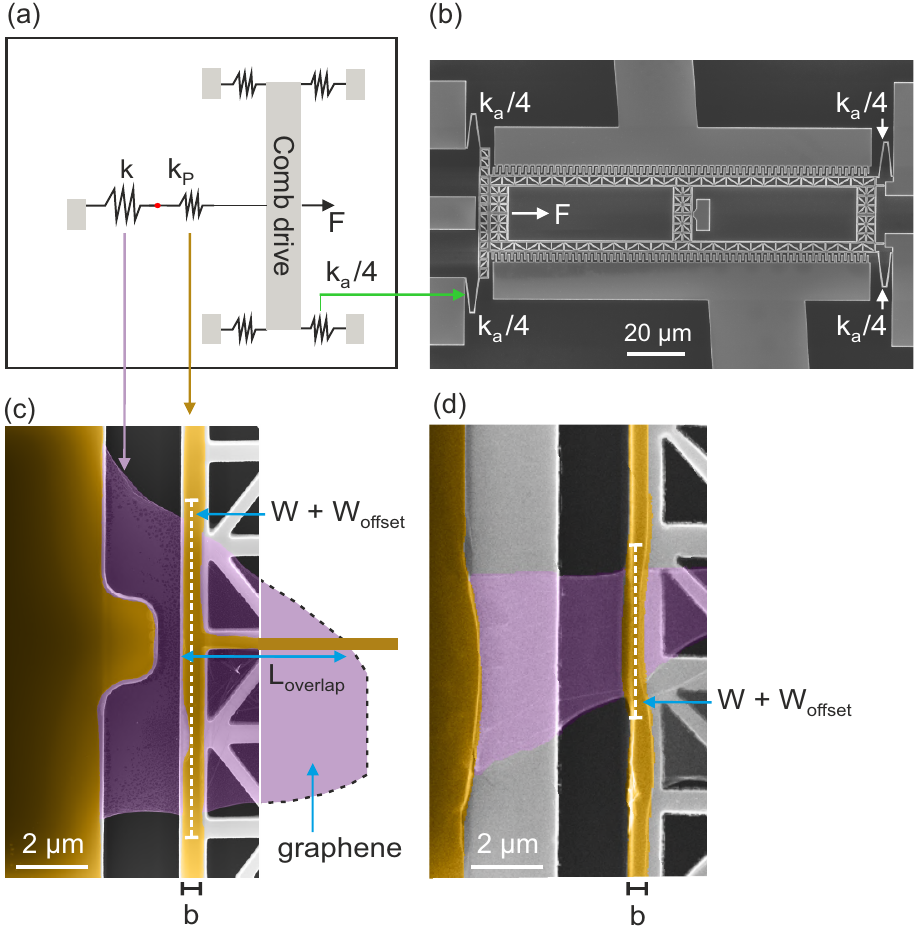}
  \caption{
    Spring configuration of the devices.
    (a) Schematic illustration of the springs in our measured samples.
    (b) The green arrow indicates the spring $k_{\text{a}}$ of the actuator.
    (c) The purple arrow indicates the spring $k$ of the graphene sheet and the dark yellow arrow indicates the spring $k_{\text{P}}$ of the clamping.
        The clamping consists of cross-linked PMMA beams along the width of the graphene flake as well as a beam along the length of the graphene flake on the actuator.
    (d) The clamping geometry of sample 1 in the main manuscript does not have the beam with length $L_{\text{overlap}}$ along along the length of the graphene flake on the actuator.
        To fully illustrate the length $L_{\text{overlap}}$, we drew, as a guide to the eye, the outline of the graphene flake beyond the SEM image.
        The white dashed lines in panels (c) and (d) indicate the part of the clamping on the silicon that is also deformed together with the graphene flake of width $W$ (see Supplementary Discussion~1).
        The length of the white dashed lines is equal to $W+W_{\text{offset}}$.
        The black lines in panels (c) and (d) indicate the width $b$ of the cross-linked PMMA beam.
    }
  \label{Sup_fig9b}
\end{figure}

\newpage

\begin{figure}[!ht]
\includegraphics[width=159mm]{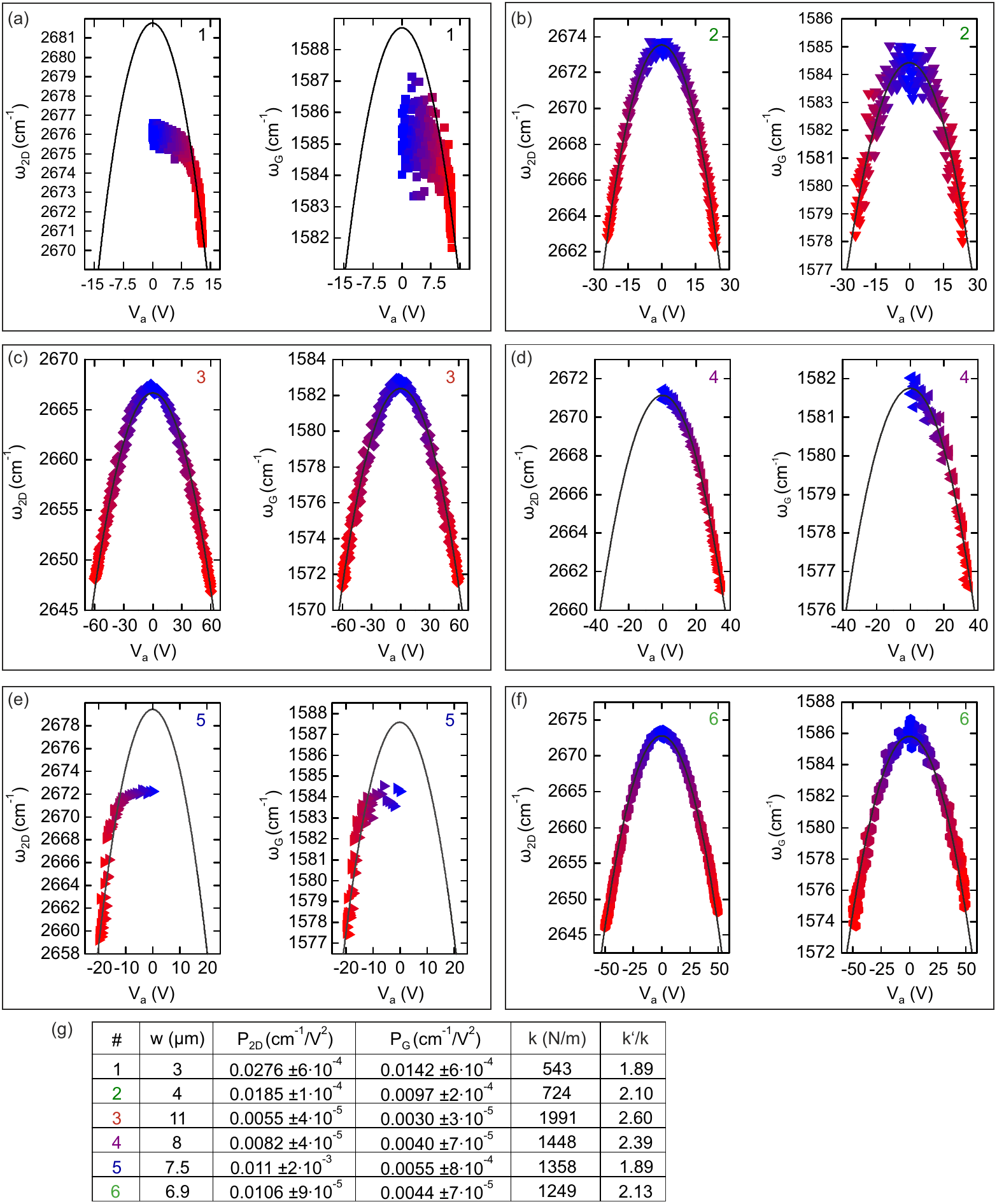}
  \caption{
    Estimation of the relative $\omega_{\text{2D(G)}}$ shift per unit strain.
    Panels (a) to (f) show the Raman 2D and G peak frequencies $\omega_{\text{2D}}$ and $\omega_{\text{G}}$ as function of the potential $V_a$ for six different samples.
    The colour of the sample numbers correspond to the one of the slope in Figure~2b of the main manuscript.
    Samples 1 and 5 are the only ones to show slack.
    The parabolic behavior agrees with the electrostatic force $F = \eta V_a^2$.
    (g) We find from fitting our data with $\Delta\omega_{\text{2D(G)}} = P_{\text{2D(G)}} V_a^2$ (see parabolas in panels (a)-(f)) and using $\partial \omega_{\text{2D}}/\partial \epsilon = -83\,\text{cm}^{\text{-1}}/\text{\%}$, the measured length $L = 2$ \textmu m and individual width $W$ as well as the estimated $k = Y_{\text{2D}} W/L$ with $Y_{\text{2D}} = 362$ N/m \cite{lee2008}, the ratio $k'/k$. As $k'/k > 1$, we must have an effective spring $k'$ that is different from the one for pristine graphene.
  }
  \label{Sup_fig6a}
\end{figure}

\newpage

\begin{figure}[!ht]
\includegraphics[width=63.1mm]{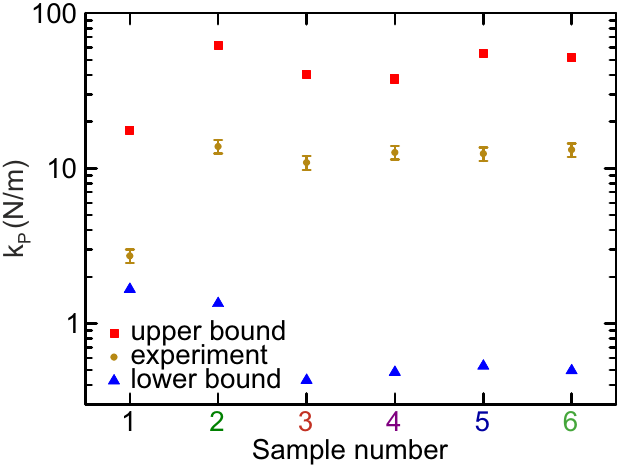}
  \caption{
    Spring constant estimation of the clamping.
    The extracted spring constant $k_{\text{P}}$ of the clamping for each different sample lies well within the bounds given by Euler-beam theory (see Supplementary Discussion~1).
    The colour of the sample numbers correspond to the one of the slope in Figure~2b of the main manuscript.
    We conclude that the clamping is elastic and acts as a spring in series to the spring $k$ of the graphene flake, which are both parallel to the spring $k_{\text{a}}$ of the actuator.
    }
  \label{Sup_fig9b}
\end{figure}

\newpage

\begin{figure}[th]
\includegraphics[width=163.1mm]{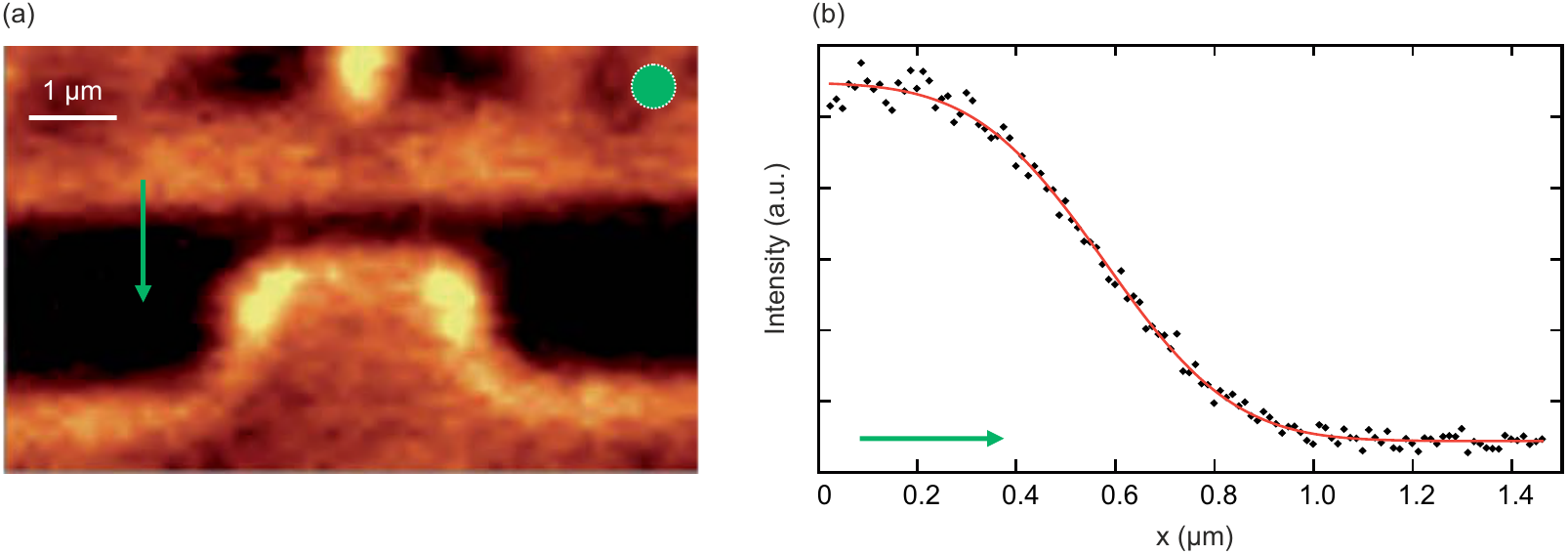}
  \caption{
	Determination of the laser spot size.
	(a) Raman map of the silicon peak intensity at 521 cm$^{\rm -1}$ of a CD actuator without a graphene flake.
	(b) Intensity of the silicon peak along the green arrow in panel (a).
        As the edge of the CD actuator is well defined and sharp (see e.g. SEM images in Supplementary Figure~2), we can understand the measured intensity profile as a convolution of a step function with a Gaussian function.
        Therefore, we can determine the laser spot size by fitting the intensity with $I(x) = I_0 \textrm{erf}(\tfrac{\sqrt{2}(x-x_0)}{r})$, in which $r$ is the radius of the Gaussian profile and $x_0$ is the position of the edge.
        The fit result, shown in red, gives us a $x_0$ of $554 \pm 4$ nm and a beam diameter of $505 \pm 10$ nm.
  }
  \label{Sup_fig5}
\end{figure}

\newpage

\begin{table}[th]
	\centering
	\label{Sup_tab1}
	\begin{tabular}{|c|c|c|c|}
		\hline
		\raisebox{8pt}{\phantom{M}} Author \raisebox{-8pt}{\phantom{M}} & & \raisebox{8pt}{\phantom{M}} $\tfrac{\partial \omega_{\text G}}{\partial\epsilon}$ $\mathrm{(cm^{-1}/\%)}$ \raisebox{-8pt}{\phantom{M}} & \raisebox{8pt}{\phantom{M}} $\tfrac{\partial \omega_{\text 2D}}{\partial\epsilon}$ $\mathrm{(cm^{-1}/\%)}$ \raisebox{-8pt}{\phantom{M}} \\
        \hline
		\raisebox{8pt}{\phantom{M}} Mohiuddin \textit{et al.}\cite{mohiuddin2009} \raisebox{-8pt}{\phantom{M}} & \raisebox{8pt}{\phantom{M}} supported \raisebox{-8pt}{\phantom{M}} & \raisebox{8pt}{\phantom{M}} -21.25 \raisebox{-8pt}{\phantom{M}} & \raisebox{8pt}{\phantom{M}} -64 \raisebox{-8pt}{\phantom{M}} \\
        \hline
		\raisebox{8pt}{\phantom{M}} Mohiuddin \textit{et al.}\cite{mohiuddin2009} \raisebox{-8pt}{\phantom{M}} & \raisebox{8pt}{\phantom{M}} suspended \raisebox{-8pt}{\phantom{M}} & \raisebox{8pt}{\phantom{M}} -27.5 \raisebox{-8pt}{\phantom{M}} & \raisebox{8pt}{\phantom{M}} -83 \raisebox{-8pt}{\phantom{M}} \\
        \hline
		\raisebox{8pt}{\phantom{M}} Yoon \textit{et al.}\cite{yoon2011} \raisebox{-8pt}{\phantom{M}} & \raisebox{8pt}{\phantom{M}} supported \raisebox{-8pt}{\phantom{M}} & \raisebox{8pt}{\phantom{M}} -23.95 \raisebox{-8pt}{\phantom{M}} & \raisebox{8pt}{\phantom{M}} A: -53.1 \raisebox{-8pt}{\phantom{M}} \\
              						        & 		    &        & \raisebox{8pt}{\phantom{M}} Z: -46.9 \raisebox{-8pt}{\phantom{M}} \\
        \hline
		\raisebox{8pt}{\phantom{M}} Frank \textit{et al.}\cite{frank2011} \raisebox{-8pt}{\phantom{M}} & \raisebox{8pt}{\phantom{M}} supported \raisebox{-8pt}{\phantom{M}} & \raisebox{8pt}{\phantom{M}} -20.5 \raisebox{-8pt}{\phantom{M}} & \raisebox{8pt}{\phantom{M}} -56.65 \raisebox{-8pt}{\phantom{M}} \\
        \hline
		\raisebox{8pt}{\phantom{M}} Polyzos \textit{et al.}\cite{polyzos2015} \raisebox{-8pt}{\phantom{M}} & \raisebox{8pt}{\phantom{M}} suspended \raisebox{-8pt}{\phantom{M}} & \raisebox{8pt}{\phantom{M}} -28 \raisebox{-8pt}{\phantom{M}} & \raisebox{8pt}{\phantom{M}} -89 \raisebox{-8pt}{\phantom{M}} \\
        \hline
		\raisebox{8pt}{\phantom{M}} Huang \textit{et al.}\cite{huang2010} \raisebox{-8pt}{\phantom{M}} & \raisebox{8pt}{\phantom{M}} supported \raisebox{-8pt}{\phantom{M}} & \raisebox{8pt}{\phantom{M}} -9.05 \raisebox{-8pt}{\phantom{M}} & \raisebox{8pt}{\phantom{M}} A: -26.1 \raisebox{-8pt}{\phantom{M}} \\
                                       		  & 		  &       & \raisebox{8pt}{\phantom{M}} Z: -23\textcolor{white}{.0} \raisebox{-8pt}{\phantom{M}} \\
        \hline
		\raisebox{8pt}{\phantom{M}} Ni \textit{et al.}\cite{ni2008} \raisebox{-8pt}{\phantom{M}} & \raisebox{8pt}{\phantom{M}} supported \raisebox{-8pt}{\phantom{M}} & \raisebox{8pt}{\phantom{M}} -14.2 \raisebox{-8pt}{\phantom{M}} & \raisebox{8pt}{\phantom{M}} -27.8 \raisebox{-8pt}{\phantom{M}} \\
        \hline
	\end{tabular}
	\caption{
      Summary of published Raman peak shifts due to strain.
      If strain was applied along the armchair and zigzag direction, this is denoted with A and Z, respectively.
      The reported values for $\tfrac{\partial \omega_{\text G}}{\partial\epsilon}$ and $\tfrac{\partial \omega_{\text 2D}}{\partial\epsilon}$ show a large spread, which can be attributed to how strain was applied to the graphene flake.
      All studies in this table used a bendable substrate except for Polyzos {\it et al.} \cite{polyzos2015}.
      This has two important consequences: (i) graphene does not exhibit its intrinsic Poisson ratio and (ii) the strain is possibly not properly transferred into the graphene flake due to the large difference between Young's modulus of the substrate and the graphene flake \cite{liu2014}.
      The latter leads to an overestimation of the induced strain and thus to an underestimation of $\tfrac{\partial \omega_{\text G}}{\partial\epsilon}$ and $\tfrac{\partial \omega_{\text 2D}}{\partial\epsilon}$.
      Therefore, we base our strain estimations on the conservative value of $\tfrac{\partial \omega_{\text 2D}}{\partial\epsilon} = -83$ $\mathrm{cm^{-1}/\%}$ reported by Mohiuddin {\it et al.} \cite{mohiuddin2009} and Polyzos {\it et al.} \cite{polyzos2015}, which is also agreeing with theory \cite{mohiuddin2009}.
    }
\end{table}

\newpage

%

\end{document}